\documentclass[reprint,showkeys,longbibliography,nofootinbib,superscriptaddress,aps,notitlepage]{revtex4-2}%

\usepackage[utf8]{inputenc}
\usepackage[dvipsnames]{xcolor}
\usepackage{mathtools}
\usepackage{amsmath}
\usepackage{amssymb}
\usepackage[colorlinks]{hyperref}
\usepackage{graphicx}
\usepackage{multirow}
\usepackage{physics}
\usepackage{siunitx}
\usepackage{xspace} 
\usepackage{lineno}
\usepackage{dcolumn}

\newcommand\myshade{80}
\colorlet{mylinkcolor}{ForestGreen}
\colorlet{mycitecolor}{Red}
\colorlet{myurlcolor}{violet}

\hypersetup{
  linkcolor  = mylinkcolor!\myshade!black,
  citecolor  = mycitecolor!\myshade!black,
  urlcolor   = myurlcolor!\myshade!black,
  colorlinks = true
}

\newcommand{\MeV}{~MeV/c$^{2}$}
\newcommand{\DMe}{~DM-e$^{-}$}
\newcommand{\mA}{$m_{A^{'}}$}
\newcommand{\qe}[1]{#1$e^{-}$}
\newcommand{\qedark}{\texttt{QEDark}}
\newcommand{\qcdark}{\texttt{QCDark}}

\usepackage{cuted}

\begin{document}
\preprint{}

\title{Daily Modulation Constraints on Light Dark Matter with DAMIC-M}

\author{K.\,Aggarwal}
\affiliation{Center for Experimental Nuclear Physics and Astrophysics, University of Washington, Seattle, WA, United States}

\author{I.\,Arnquist}
\affiliation{Pacific Northwest National Laboratory (PNNL), Richland, WA, United States} 

\author{N.\,Avalos}
\affiliation{Laboratoire de physique nucl\'{e}aire et des hautes \'{e}nergies (LPNHE), Sorbonne Universit\'{e}, Universit\'{e} Paris Cit\'{e}, CNRS/IN2P3, Paris, France}

\author{X.\,Bertou}
\affiliation{CNRS/IN2P3, IJCLab, Universit\'{e} Paris-Saclay, Orsay, France}
\affiliation{Laboratoire de physique nucl\'{e}aire et des hautes \'{e}nergies (LPNHE), Sorbonne Universit\'{e}, Universit\'{e} Paris Cit\'{e}, CNRS/IN2P3, Paris, France}

\author{N.\,Castell\'{o}-Mor}
\affiliation{Instituto de F\'{i}sica de Cantabria (IFCA), CSIC - Universidad de Cantabria, Santander, Spain}

\author{C.\,Centeno-Lorca}
\affiliation{Instituto de F\'{i}sica de Cantabria (IFCA), CSIC - Universidad de Cantabria, Santander, Spain}

\author{A.E.\,Chavarria}
\affiliation{Center for Experimental Nuclear Physics and Astrophysics, University of Washington, Seattle, WA, United States}

\author{J.\,Cuevas-Zepeda}
\affiliation{Kavli Institute for Cosmological Physics and The Enrico Fermi Institute, The University of Chicago, Chicago, IL, United States}

\author{A.\,Dastgheibi-Fard}
\affiliation{LPSC LSM, CNRS/IN2P3, Universit\'{e} Grenoble-Alpes, Grenoble, France}

\author{C.\,De Dominicis}
\affiliation{Laboratoire de physique nucl\'{e}aire et des hautes \'{e}nergies (LPNHE), Sorbonne Universit\'{e}, Universit\'{e} Paris Cit\'{e}, CNRS/IN2P3, Paris, France}

\author{O.\,Deligny}
\affiliation{CNRS/IN2P3, IJCLab, Universit\'{e} Paris-Saclay, Orsay, France}

\author{J.\,Duarte-Campderros}
\affiliation{Instituto de F\'{i}sica de Cantabria (IFCA), CSIC - Universidad de Cantabria, Santander, Spain}

\author{E.\,Estrada}
\affiliation{Centro At\'{o}mico Bariloche and Instituto Balseiro, Comisi\'{o}n Nacional de Energ\'{i}a At\'{o}mica (CNEA), Consejo Nacional de Investigaciones Cient\'{i}ficas y T\'{e}cnicas (CONICET), Universidad Nacional de Cuyo (UNCUYO), San Carlos de Bariloche, Argentina}

\author{R.\,Ga\"{i}or}
\affiliation{Laboratoire de physique nucl\'{e}aire et des hautes \'{e}nergies (LPNHE), Sorbonne Universit\'{e}, Universit\'{e} Paris Cit\'{e}, CNRS/IN2P3, Paris, France}

\author{E.-L.~Gkougkousis}
\affiliation{Universit\"{a}t Z\"{u}rich Physik Institut, Z\"{u}rich, Switzerland}

\author{T.\,Hossbach}
\affiliation{Pacific Northwest National Laboratory (PNNL), Richland, WA, United States} 

\author{L.\,Iddir}
\affiliation{Laboratoire de physique nucl\'{e}aire et des hautes \'{e}nergies (LPNHE), Sorbonne Universit\'{e}, Universit\'{e} Paris Cit\'{e}, CNRS/IN2P3, Paris, France}

\author{B.~J.~Kavanagh}
\affiliation{Instituto de F\'{i}sica de Cantabria (IFCA), CSIC - Universidad de Cantabria, Santander, Spain}

\author{B.\,Kilminster}
\affiliation{Universit\"{a}t Z\"{u}rich Physik Institut, Z\"{u}rich, Switzerland}


\author{I.\,Lawson}
\affiliation{SNOLAB, Lively, ON, Canada }

\author{A.\,Letessier-Selvon}
\affiliation{Laboratoire de physique nucl\'{e}aire et des hautes \'{e}nergies (LPNHE), Sorbonne Universit\'{e}, Universit\'{e} Paris Cit\'{e}, CNRS/IN2P3, Paris, France}

\author{H.\,Lin}
\affiliation{Department of Physics and Astronomy, Johns Hopkins University, Baltimore, MD, United States}

\author{P.\,Loaiza}
\affiliation{CNRS/IN2P3, IJCLab, Universit\'{e} Paris-Saclay, Orsay, France}

\author{A.\,Lopez-Virto}
\affiliation{Instituto de F\'{i}sica de Cantabria (IFCA), CSIC - Universidad de Cantabria, Santander, Spain}

\author{R.\,Lou}
\affiliation{Kavli Institute for Cosmological Physics and The Enrico Fermi Institute, The University of Chicago, Chicago, IL, United States}

\author{H. Lumengo-Kidimbu}
\affiliation{Laboratoire de physique nucl\'{e}aire et des hautes \'{e}nergies (LPNHE), Sorbonne Universit\'{e}, Universit\'{e} Paris Cit\'{e}, CNRS/IN2P3, Paris, France}

\author{K.~J.\,McGuire}
\affiliation{Center for Experimental Nuclear Physics and Astrophysics, University of Washington, Seattle, WA, United States}

\author{S.\,Munagavalasa}
\affiliation{Kavli Institute for Cosmological Physics and The Enrico Fermi Institute, The University of Chicago, Chicago, IL, United States}

\author{J.\,Noonan}
\affiliation{Kavli Institute for Cosmological Physics and The Enrico Fermi Institute, The University of Chicago, Chicago, IL, United States}

\author{D.\,Norcini}
\affiliation{Department of Physics and Astronomy, Johns Hopkins University, Baltimore, MD, United States}

\author{S.\,Paul}
\affiliation{Kavli Institute for Cosmological Physics and The Enrico Fermi Institute, The University of Chicago, Chicago, IL, United States}

\author{P.\,Privitera}
\affiliation{Kavli Institute for Cosmological Physics and The Enrico Fermi Institute, The University of Chicago, Chicago, IL, United States}
\affiliation{Laboratoire de physique nucl\'{e}aire et des hautes \'{e}nergies (LPNHE), Sorbonne Universit\'{e}, Universit\'{e} Paris Cit\'{e}, CNRS/IN2P3, Paris, France}
 \affiliation{Laboratoire de Physique, \'{E}cole Normale Sup\'{e}rieure, Sorbonne Universit\'{e}, Universit\'{e} Paris Cit\'{e}, CNRS/IN2P3, Paris, France}

\author{P.\,Robmann}
\affiliation{Universit\"{a}t Z\"{u}rich Physik Institut, Z\"{u}rich, Switzerland}

\author{B.\,Roach}
\affiliation{Kavli Institute for Cosmological Physics and The Enrico Fermi Institute, The University of Chicago, Chicago, IL, United States}

\author{D.\,Rosenmerkel}
\affiliation{Department of Physics and Astronomy, Johns Hopkins University, Baltimore, MD, United States}

\author{M.\,Settimo}
\affiliation{SUBATECH, Nantes Universit\'{e}, IMT Atlantique, CNRS/IN2P3, Nantes, France}

\author{R.\,Smida}
\affiliation{Kavli Institute for Cosmological Physics and The Enrico Fermi Institute, The University of Chicago, Chicago, IL, United States}

\author{M.\,Traina}
\affiliation{Instituto de F\'{i}sica de Cantabria (IFCA), CSIC - Universidad de Cantabria, Santander, Spain}

\author{R.\,Vilar}
\affiliation{Instituto de F\'{i}sica de Cantabria (IFCA), CSIC - Universidad de Cantabria, Santander, Spain}

\author{R.\,Yajur}
\affiliation{Kavli Institute for Cosmological Physics and The Enrico Fermi Institute, The University of Chicago, Chicago, IL, United States}

\author{D.\,Venegas-Vargas}
\affiliation{Department of Physics and Astronomy, Johns Hopkins University, Baltimore, MD, United States}

\author{C.\,Zhu}
\affiliation{Department of Physics and Astronomy, Johns Hopkins University, Baltimore, MD, United States}

\author{Y.\,Zhu}
\affiliation{Laboratoire de physique nucl\'{e}aire et des hautes \'{e}nergies (LPNHE), Sorbonne Universit\'{e}, Universit\'{e} Paris Cit\'{e}, CNRS/IN2P3, Paris, France}

\collaboration{DAMIC-M Collaboration}

\date{\today} 

\begin{abstract}
\noindent  The flux of Hidden Sector particles from the Galactic halo reaching an underground detector can be significantly attenuated by interactions within the Earth for sufficiently large scattering cross-sections. This attenuation gives rise to a characteristic daily modulation in the detection rate, due to Earth's rotation. We present results from a search for such a modulation using a 1.257 kg-day dataset collected with the DAMIC-M Low Background Chamber.
A model-independent analysis reveals no significant modulation in the \qe{1} event rate over periods from 1 to 48 h, highlighting the excellent temporal stability of the detector. In a complementary model-dependent analysis, we target the expected daily modulation signature of Hidden Sector particles, with masses in the range $[0.53, 2]$~$\mathrm{MeV/c^2}$, interacting with electrons via a dark photon mediator. By leveraging the expected temporal evolution of the signal, we set improved constraints on Dark Matter masses below 1.2~$\mathrm{MeV/c^2}$, surpassing our previous limits.
 
\end{abstract}

\maketitle


\section{Introduction} 
\label{sec:introduction}

Dark Matter (DM) is one of the largest outstanding mysteries in modern cosmology, astrophysics and particle physics~\cite{Bertone:2004pz}. One possibility is that DM takes the form of light MeV-GeV scale particles, which reside in Dark Sectors containing new particles, symmetries and forces~\cite{Boehm:2003ha,Pospelov:2007mp,Hooper:2008im,Chu:2011be,Knapen:2017xzo}. 

Although these particles are only weakly coupled to the Standard Model, they are expected to scatter off nuclei and electrons in targets made of ordinary matter.
Due to their low masses, these light Dark Sector particles cannot transfer enough energy to nuclei for detection via traditional nuclear recoil-based direct detection methods~\cite{Drukier:1983gj,Goodman:1984dc,Drukier:1986tm,Billard:2021uyg}. Instead, electron recoil-based experiments with semiconductor targets, in particular silicon Charge-Coupled Devices (CCDs), have become the leading alternative~\cite{Essig:2011nj}. 

To date, the sensitivity of these experiments at $\mathcal{O}$(1 MeV) masses has been limited by background from single-electron events that are challenging to distinguish from potential DM signals~\cite{DAMIC-M:2023FC,Adari:2022}. Adari
These backgrounds have two main components~\cite{SEEsensei2022}: an exposure-dependent one, which includes CCD thermal dark current, external infrared photons, and Cherenkov radiation from particle interactions in and around the CCDs; and an exposure-independent component inherent to the CCD design and operating parameters, including spurious charge from charge transfer inefficiency.
For simplicity, we will refer to these backgrounds as dark current throughout this work.
A DM signal would introduce an additional exposure-dependent component to the single-electron rate and, unlike background events, is expected to exhibit a modulation on a sidereal timescale. We can exploit such a signature to improve the sensitivity of our DM search.

For MeV-scale DM, when the cross sections exceed $\sim10^{-36}$~cm$^{2}$ for DM-proton and $\sim10^{-33}$~cm$^{2}$ for DM-electron scattering,
the DM mean free path becomes comparable to the Earth's radius~\cite{Emken:2017qmp,Emken:2019tni}. In this regime, Earth-scattering effects can significantly reshape the DM speed distribution at the detector~\cite{Gould:1988eq,Collar:1993ss,Hasenbalg:1997hs}. 
Although the total event rate may decrease at very large cross sections, a unique and robust signature emerges: a sidereal daily modulation in the recoil rate, induced by the rotation of the Earth relative to the incoming Galactic DM flux~\cite{Kouvaris:2014lpa,Kavanagh:2016pyr,Emken:2017qmp,Emken:2019tni}. Such a modulation provides a powerful discriminator against un-modulated backgrounds, especially in the single-electron regime, where conventional spectral analyses are limited.

The DAMIC-M experiment~\cite{Privitera:2024}, which employs CCDs as sensitive targets, reported the first dedicated search for daily modulation induced by Earth-scattering effects in DM-electron interactions~\cite{DAMIC-M:2023DM}, leveraging the detector’s excellent stability.
In this work, we build upon that analysis by performing a follow-up search for daily modulation using 
an improved dataset with a tenfold increase in exposure ($\sim1.3$ kg-day) and a fiftyfold reduction in the single-electron background rate.
A previous search for events with two or more electrons in this dataset placed the most stringent constraints to date on DM-electron scattering over a broad mass range, and, notably, was the first to exclude a well-motivated region of parameter space for freeze-in production scenarios~\cite{DAMIC-M:2025}.

The combination of reduced backgrounds, increased exposure, and sensitivity to time-dependent signatures enables the exploration of new parameter space for sub-MeV-scale Hidden-Sector DM, particularly in scenarios involving ultralight mediators. Our results underscore both the enhanced performance of the DAMIC-M prototype~\cite{LBC:2024} and the role of time-domain techniques in the broader search for light DM.


\section{Setup and data} 
\label{sec:setup}

This work is based on the same dataset used in our previous DM search~\cite{DAMIC-M:2025}. A brief description of the experimental setup and dataset follows.

Data is acquired with the DAMIC-M prototype detector, known as the Low Background Chamber (LBC)~\cite{LBC:2024}, operating at the Modane Underground Laboratory (LSM). 
The LBC houses two DAMIC-M prototype CCD modules (Module 1 and Module 2), each comprising four high-resistivity ($>$10 k$\Omega$cm) $n$-type silicon CCDs~\cite{Holland:2002zz,Holland:2003zz,Holland:2009zz} mounted on a silicon pitch adapter with traces for clocks, bias, and video signals. For data acquisition, we used the final version of the DAMIC-M CCD controller and front-end electronics. The active region of each CCD is a rectangular array of $6144 \times 1536$ pixels, each of area $15 \times 15$~$\mu$m$^2$, with a thickness of $670$ $\mu$m, and a mass of approximately 3.3~g. 
The CCDs are enclosed in a high-purity copper box with lids that were electroformed underground, providing shielding against infrared radiation. 
The devices inside the LBC vacuum cryostat are cooled to a temperature of $\sim$130 K via a cryocooler and maintained at a pressure of $\sim5\times 10^{-6}$ mbar. Additional shielding from environmental radiation is provided by layers of lead and polyethylene surrounding the CCD box and the cryostat. Except for the CCD box and the copper cold fingers, which transfer cooling from the cryocooler to the devices, the rest of the setup remains at room temperature.

The CCDs are fully depleted by applying a substrate bias voltage of 45~V. Charge carriers, which are generated by ionizing radiation within the silicon bulk, drift perpendicularly to the $x$–$y$ plane of the pixel array under the influence of the external electric field. While drifting, thermal diffusion causes the charge to spread and distribute between adjacent pixels with variance $\sigma^2_{xy}$~\cite{DAMICsnolab}.
During readout, the charge is transferred across the pixel array through voltage clocking of the three-phase pixel gates, which also permits pixel binning, where charge from multiple pixels is summed before being read out\footnote{An ${\it n \times m}$ binning refers to summing the charge from {\it n} pixels along the horizontal axis and {\it m} along the vertical axis.}. The charge is read out serially by an amplifier located at the end of the serial register, which can be seen as the first row of the pixel array. In the DAMIC-M modules, each CCD is readout through one amplifier.
The skipper amplifier~\cite{skipper,Chandler1990zz,Tiffenberg:2017aac} enables multiple non-destructive charge measurements (NDCMs) per pixel. By averaging $N_\mathrm{NDCM}$ such measurements, sub-electron resolution is achieved, improving as $1/\sqrt{N_\mathrm{NDCM}}$. All data reported in this work is acquired using $N_\mathrm{NDCM} = 500$, yielding a pixel charge resolution $\rm{\sigma_{ch}}$ of approximately 0.16~$e^-$, and ${1\times100}$ binning. Unless otherwise stated, the term pixel will henceforth refer to a binned pixel.

For this analysis, we use $\sim84$~days of data taken between October 2024 and January 2025. 
Data is acquired continuously and is recorded in images of $6300\times16$ pixels\footnote{Pixels in the image beyond the physical dimensions of the CCDs, commonly referred to as ``overscan'', are exposed for a much shorter time than the other pixels. The overscan region, while not used for the analysis, provides valuable information about data quality and dark current in the serial register.}. 
Each CCD pixel is exposed for a time approximately equal to the image readout time of 1668~s, plus 6~s of waiting time between images.
The data-set is divided into two parts: D1 ($\sim 7$~days) is used to develop the masking and data selection procedures, which are subsequently applied in a blind manner to D2 ($\sim 77$~days), used for the DM search.
Two CCDs (2-A and 2-C) belonging to Module 2 are excluded from the analysis. The first features a defect at the edge of the serial register that results in an excess of pixels with charge  $>$1$e^-$. The second shows a large noise, preventing single electron resolution.
Due to increased noise injected into Module 2 by the malfunctioning CCD 2-C, both CCDs 2-C and 2-B---sharing the same amplifier voltage lines---were switched off 
\footnote{Here, ``switching off" refers to under-volting the drain of the output amplifier.} during the final $\sim 47$ days of D2. Module 2 was fully shut down following a brief data acquisition incident that occurred $\sim 20$ days before the end of D2. Approximately 15 hours before switching off CCDs 2-C and 2-B, the cryocooler power was raised to ensure adequate cooling of the devices.

A brief summary of the image data reduction and pixel selection criteria is provided here, while a detailed description can be found in the End Matter of Ref.~\cite{DAMIC-M:2025}. Note that the selection procedure was optimized for our previous analysis~\cite{DAMIC-M:2025}, during which the temporal evolution of the data, relevant to the present work, was kept blinded.
Each CCD image consists of an array of pixel charges in ADUs (analog-to-digital units). The image pedestal is first calculated and subtracted row by row, and then column by column, so that the value of pixels without charge is centered at 0. 
CCD images are calibrated from ADUs to electron units from the distance between the first two peaks in the pixel charge distribution, i.e., the \qe{0} and \qe{1} peaks. 
Defects in the CCDs, appearing in the images as ``hot" columns~\cite{2001sccd.book.....J}, are identified by their high rate of \qe{1} pixels, and rejected. 
We exclude a high dark current region in CCD 1-C for columns larger than 5315, where a prominent hot column is present. 
Clusters of adjacent pixels with total charge $\rm{q>5e^- + 3\,\sigma_{ch}}$ are masked since their energy is too high to be produced by sub-GeV DM interactions with electrons. To reject crosstalk and correlated noise, pixels belonging to a cluster are masked in all CCDs of the same module.  Pixels with high variance across the $N_\mathrm{NDCM}$ measurements are also excluded from the analysis together with pixels with charge positively or negatively correlated between different CCDs of the same module.
To eliminate trailing charge released from traps in the CCD active area or in the serial register, we reject 100 pixels above any pixel with charge $>$\qe{100} as well as all pixels in its row. Finally, we reject columns and rows with an anomalously large number of  pixels with charge $\ge3.75\rm{\sigma_{ch}}$ or with more than one pixel with charge $\ge(1 e^-+3.75\rm{\sigma_{ch})}$. 
With these selection criteria, about 95\% of the pixels in the active area of the CCDs are retained for the analysis, except for CCD 1-C, where only about 85\% of the pixels are kept due to the prominent hot column mentioned earlier.

The resulting integrated exposures are 0.139 kg-days for D1 and 1.257 kg-days for D2.

\section{Daily modulation analysis}
\label{sec:dma}

In this section, we investigate the presence of periodic signals in our dataset, with particular attention to a potential daily modulation.

\begin{figure}
    \centering
    \includegraphics[width=\linewidth]{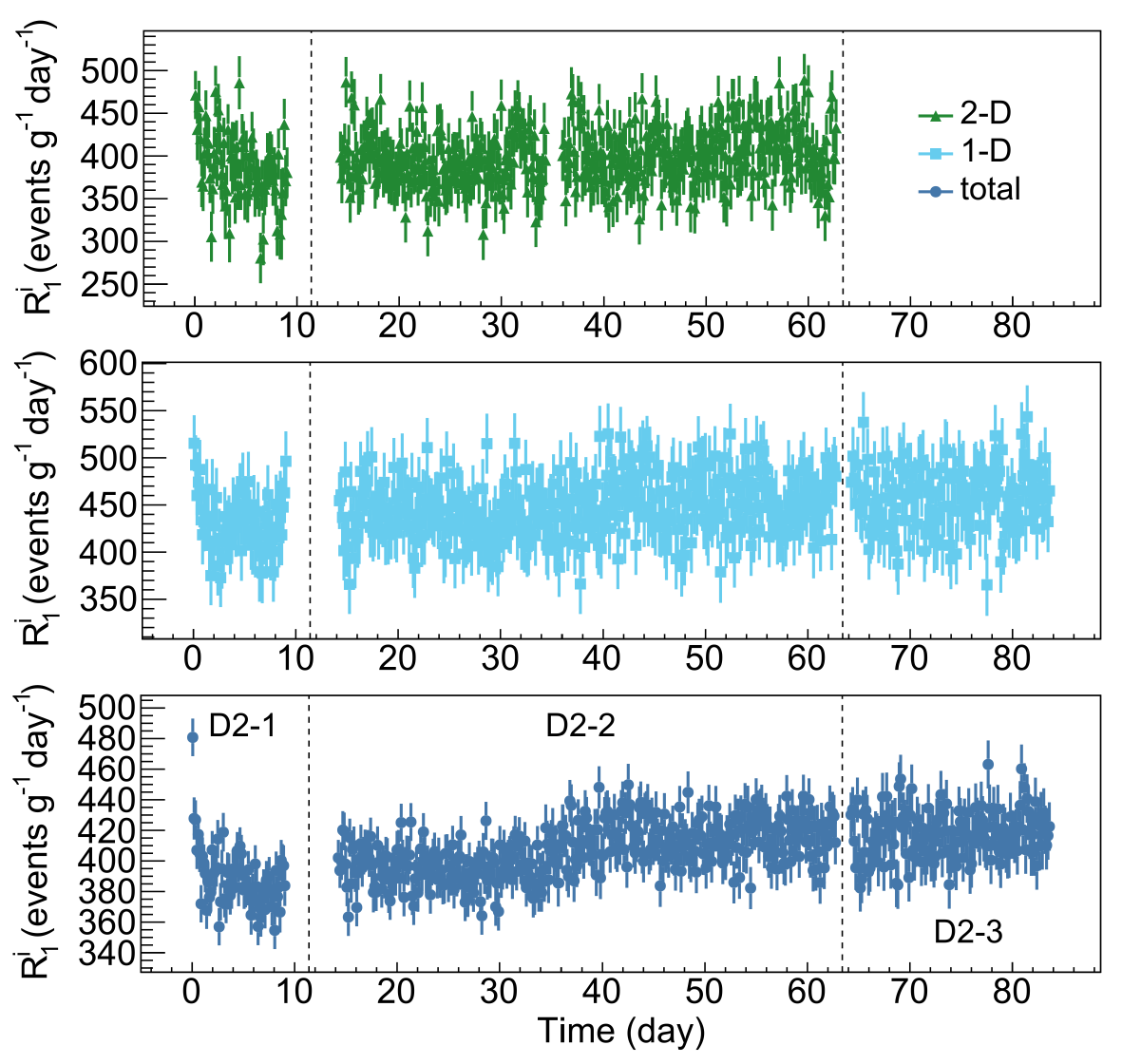}
    \caption{$R_1^i$ as a function of time for CCD 1-D (middle), CCD 2-D (top) during D2. The temporal evolution of the total $R_1^i$, derived summing the contributions of all the CCDs, is also shown (bottom). The dashed lines indicate the boundaries between the different segments of the dataset, labeled D2-1, D2-2, and D2-3. To improve visibility, we averaged approximately seven consecutive values of $R_1^i$ within each portion of the dataset between the temporal gaps. The error on the mean is shown. During the temporal gap between D2-1 and D2-2, a dedicated run was acquired to characterize defects in the CCDs. The gap between D2-2 and D2-3 is due to a brief data acquisition incident, following which Module 2 was shut down. An additional smaller gap is visible for CCD 2-D during D2-2, which corresponds to the period of time when the CCD 2-C noise increased exponentially, before turning the malfunctioning CCD off. }
    \label{fig:R1}
\end{figure}

For each CCD image $i$\footnote{Note that image $i$ for each CCD is acquired at the same time across all CCDs.}, we count the number of \qe{1} pixels ($N_{1}^{i}$) selecting the pixels with charge $\ge 4.4 \rm{\sigma_{ch}}$ and $\le 1e^- + 3.125 \rm{\sigma_{ch}}$. These cuts are chosen to maximize the signal of \qe{1} events over the leakage of \qe{0} pixels due to charge resolution.
The resolution $\rm{\sigma_{ch}}$ is estimated for every image, by fitting the \qe{0} and \qe{1} peaks of the pixel charge distribution with the sum of two Gaussian functions, with the position of the two peaks fixed.
We find the value of $N_{1}^{i}$ to be $\sim25$ for each image.
The \qe{1} rates, $R_1^i$, are obtained dividing $N_{1}^{i}$ by the effective pixel exposure time ($t_\mathrm{exp}=0.019206$ days) and target mass ($m^i_t$) after pixel selection: $R_1^i \sim 402~e^-\rm{g^{-1} day^{-1}}$. Notably, $R_1^i$ is reduced by a factor of $\sim$50 relative to Ref.~\cite{DAMIC-M:2023FC}, due to improvements in the readout electronics and in the light-tightening of the CCD box. $R_1^i$ is  measured approximately every 28 minutes over 77 days in D2.  As an example, we show in Fig.~\ref{fig:R1} the time variation of $R_1^i$ for CCD 1-D and CCD 2-D during D2. 

In our dataset, the \qe{1} rate shows a clear correlation with the system temperature measured outside the CCD box, at the level of the lead shield inside the cryostat, which is monitored continuously during data taking. We attribute this correlation to infrared photons emitted by room‑temperature components surrounding the CCD box that reach the CCDs. 
For each CCD, we model the dependence of the \qe{1} rate on temperature as:
\begin{equation}
    r^i = r_0 + \alpha_T T_i,
    \label{eq:T}
\end{equation}
where $T_i$ is the average temperature during the acquisition of an image, $r_0$ is the baseline rate, and $\alpha_T$ quantifies the linear dependence on temperature. 

We then perform an unweighted least-squares fit to the data to determine the best-fit parameters of this model. 
This procedure is applied separately to each segment of the dataset (D2-1, D2-2, and D2-3 in Fig.\ref{fig:R1}), for which the parameter $r_0$ is not necessarily expected to be the same. 
We fix $\alpha_T$ to be the weighted average of the values from all the segments, as the temperature dependence is assumed to be consistent across all data. 
We confirmed that the fit results are compatible with those obtained when $\alpha_T$ is left free.
The $\alpha_T$ parameters are consistent across the different CCDs, as shown in Table~\ref{tab:alpha_t} .
\begin{table}[]
    \centering
    \begin{tabular}{c c c}
    \hline\hline 
         \multicolumn{3}{c}{$\alpha_T$ ($e^-\rm{g}^{-1}\rm{day}^{-1}\rm{K}^{-1}$)} \\ \hline
       CCD & Module 1 & Module 2 \\ \hline 
         A & 30.9 $\pm$ 5.0 & \\
         B & 31.8 $\pm$ 4.7 &  14.9 $\pm$ 9.2\\
         C & 41.1 $\pm$ 5.2 & \\
         D & 33.1 $\pm$ 5.0 & 21.9 $\pm$ 5.2\\ \hline\hline
    \end{tabular}
    \caption{Weighted average, and associated error, of the $\alpha_T$ parameters obtained for the different D2 segments for each CCD.}
    \label{tab:alpha_t}
\end{table}

We use the fit result to estimate the expected mean number of pixels with \qe{1} in each image, $\mathcal{N}_{1}^{i} = r^i \times m^i_t \times t_{\text{exp}}$. We assign to each observed value $N_{1}^{i}$ a Poisson uncertainty $\sigma_{N_{1}}^{i}=\sqrt{\mathcal{N}_{1}^{i}}$.
The uncertainties on the rate, $\sigma_{R_{1}}^{i}$, are obtained by dividing $\sigma_{N_{1}}^{i}$ by the target mass and exposure time. 

We also compute the total $R_1^i$ by summing the individual CCD $N_{1}^{i}$ contributions and dividing the result by the total target mass and exposure time (see Fig.~\ref{fig:R1}, bottom). The associated Poisson uncertainties are evaluated as before.

\begin{figure}
    \centering
    \includegraphics[width=\linewidth]{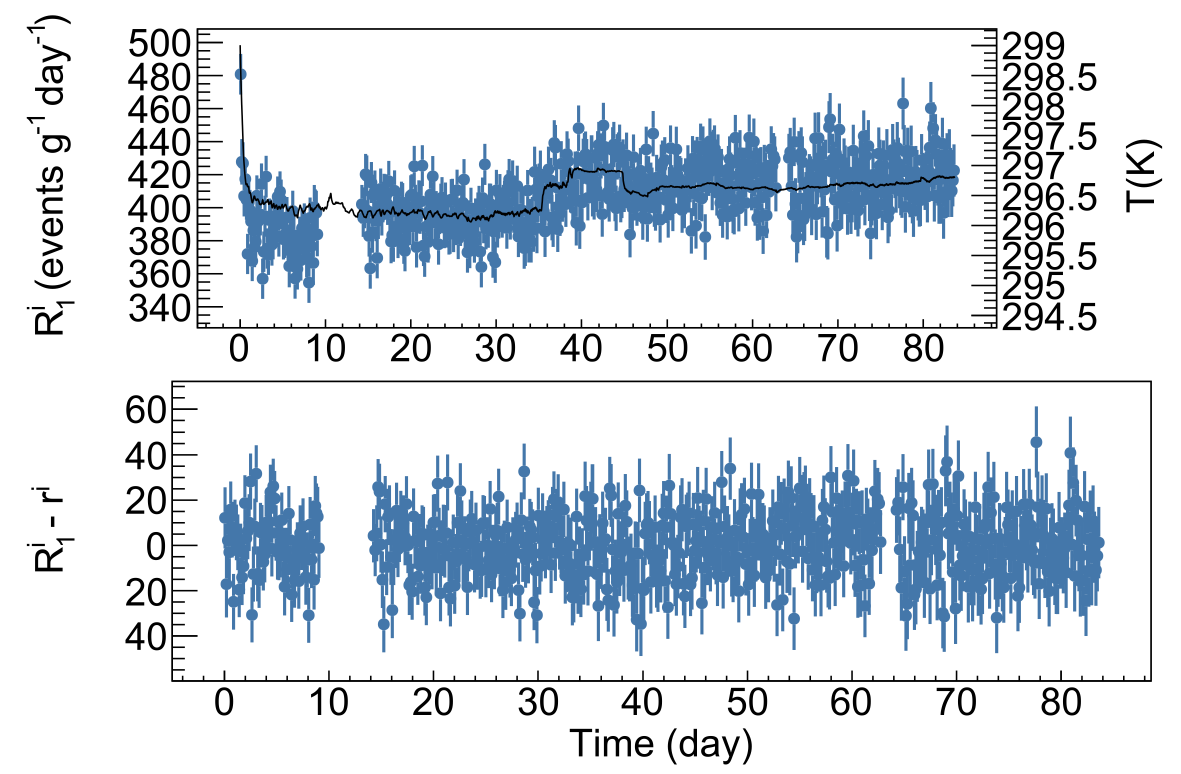}
    \caption{ Top: the temporal evolution of the total $R_1^i$ (blue dots) together with that of the system temperature (black line), which is measured at the level of the lead shield inside the cryostat, surrounding the CCD box. Bottom: the temporal evolution of the total $R_1^i$ after subtracting $r^i(T_i)$. }
    \label{fig:R1_T}
\end{figure}

 In the top panel of Fig.~\ref{fig:R1_T}, the temporal evolution of the total \qe{1} rate is presented alongside the system temperature. The bottom panel shows the total rate after subtracting $r^i$.

We analyze the temporal evolution of $R_1^i$ with two different approaches.
The model-independent approach evaluates detector stability and tests for the presence of a periodic signal.
The model-dependent approach specifically targets the expected daily modulation signature of leptophilic DM particles from the Galactic halo.


\subsection{Model-independent search for periodic signals}
\label{sec:modelindependent}

To detect significant periodic signals in our dataset, we employed the Lomb-Scargle periodogram method \cite{Lomb, Scargle, VanderPlas_2018}, which is well suited for analyzing unevenly spaced observations.

We evaluate the periodogram on $R_1^i(t_i)-r^i$ for 9832 (4042 and 7376) 
equally spaced frequencies between $1/48$~h$^{-1}$ and 1 h$^{-1}$ for Module 1 (for CCD 2-B and 2-D, respectively).
The frequency spacing is set to $1/(5 t_{tot})$, with $t_{tot}$ being the time span between the first and last points of the dataset, which is sufficiently small to resolve the peaks in the periodogram.
For every frequency $f$, we use a $\chi^2$ approach to fit the time series with a sinusoidal model centered at the weighted mean of the data: 
\begin{equation}
    y(\vec\varphi; f, t) = \varphi_0 + \varphi_1 \sin(2\pi t f ) +  \varphi_2 \cos(2\pi t f )\,,
\label{eq:y}
\end{equation}
where $\vec\varphi=(\varphi_0,\varphi_1,\varphi_2)$ is a vector containing the fit parameters, and $\varphi_0$ is the offset with respect to the data weighted mean. We call $\hat{\chi}^2(f)$ the minimum $\chi^2$ value found for the best fit model at each frequency. 
We then use the ``PSD normalization'' to construct the periodogram peak amplitude $P_{LS}$, which is comparable to the standard Fourier power spectral density (PSD) \cite{VanderPlas_2018}: 
\begin{equation}
    P_{LS}(f) = \frac{1}{2} (\hat{\chi_0}^2 - \hat{\chi}^2(f)) \, , 
\end{equation}
where  $\hat{\chi_0}^2$ is the best-fit $\chi^2$ considering a constant function model.

We assess the significance of the peaks in the periodogram by computing, for each CCD, the false-alarm probability (FAP), \emph{i.e.}, the probability that a dataset with no periodic component would produce a peak of magnitude equal or greater than a given $P_{LS}$ value. This is done using a bootstrap technique, in which we resample each CCD dataset by drawing randomly with replacement from the observed rate values while keeping the original time coordinates fixed.
For each resampled dataset, we calculate the periodogram and record its maximum amplitude. We use $10^4$ resamplings and estimate the $\mathrm{FAP}=0.159$ and $0.023$ levels as the 0.841 and 0.977 quantiles of the maximum amplitude distribution (corresponding to the quantiles at $1\sigma$ and $2\sigma$ for a normal distribution, respectively). 

\begin{figure}
    \centering
    \includegraphics[width=\linewidth]{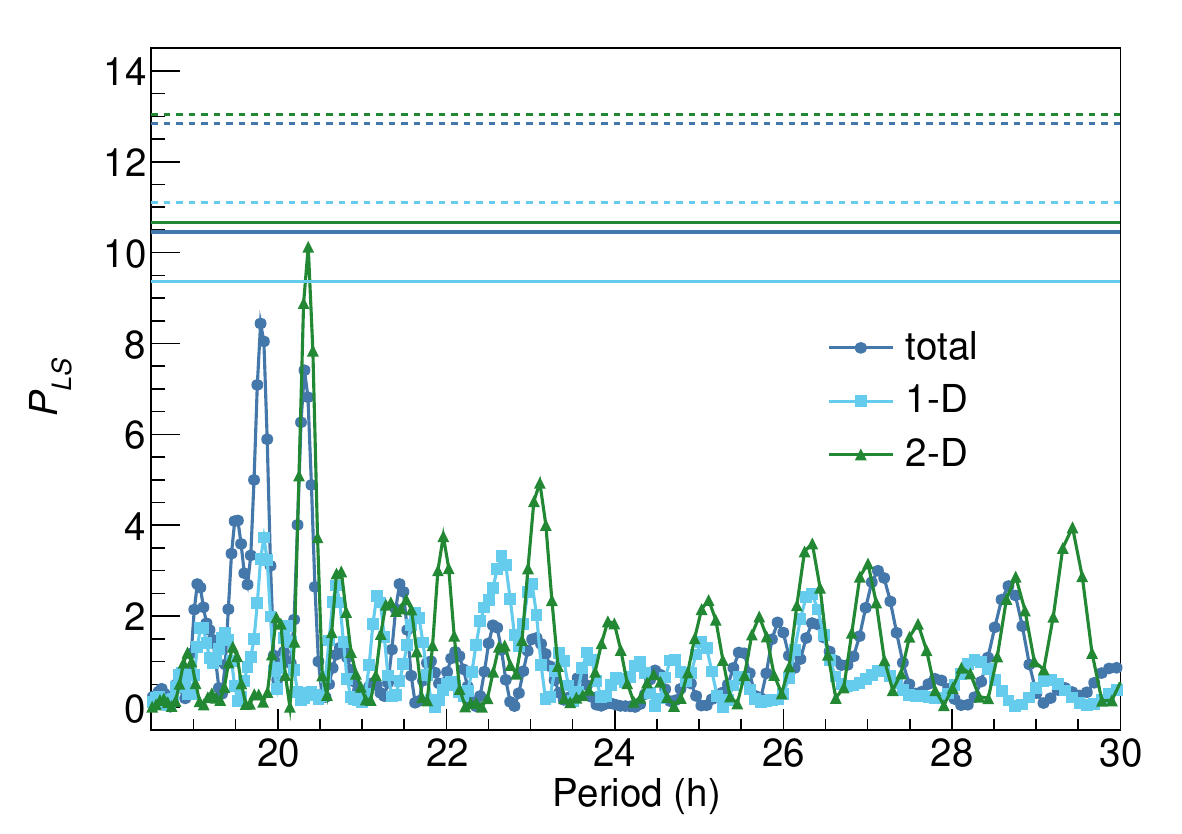}
    \caption{Lomb-Scargle periodograms calculated on the CCD 1-D, CCD 2-D, and total \qe{1} rates during D2. The continuous and dashed horizontal lines represent the $\mathrm{FAP}=0.159$ and $0.023$ levels, respectively. For clarity, we display the periodograms only over the period range of $18.5-30$ h, although they were computed over the full range of $1-48$ h.}
    \label{fig:PLS}
\end{figure}

None of the CCDs exhibit statistically significant peaks in their periodograms, demonstrating the absence of a periodic signal. We show in Fig. \ref{fig:PLS} the periodograms for two example CCDs, CCD 1-D and CCD 2-D, and for all CCDs combined using the total \qe{1} rate.

As a cross-check, we repeated the model-independent analysis using the approach from our previous publication~\cite{DAMIC-M:2023DM}. Specifically, we performed a likelihood-ratio test to evaluate the presence of periodic components in the dataset. The results are in excellent agreement with those obtained using the Lomb-Scargle periodogram.

\subsection{Model-dependent search for daily modulation of DM-e scattering signals}
\label{sec:dmescattering}

In this section, we follow our previous model-dependent approach for constraining \DMe{} scattering~\cite{DAMIC-M:2023DM}, which exploits the expected daily modulation due to DM interactions within the Earth. 

Let $F_1^i$ denote the expected number of \qe{1} events in a CCD image $i$, arising from unknown backgrounds $B^i$ and a hypothesized signal $S^i$: 
\begin{equation}
\begin{aligned}
     F_1^i(t_i , T_i, \mathbf{b}, \mathbf{s} ) & = N^i_\mathrm{pix} \sum\limits_{n=0}^{2} \sum\limits_{m+k=n}  B^i(m | \mathbf{b},T_i) \, S^i(k|\mathbf{s},t_i)\epsilon^i_n 
     \,,
\end{aligned}     
\end{equation}

where $N^i_\mathrm{pix}$ is the number of used pixels in each image, $t_i$ is the image start time, $\epsilon^i_n$ is the probability for a pixel with  \qe{n} to be selected as \qe{1} pixel due to charge resolution,  and $\mathbf{b}=\{\eta_0, c^l, \alpha_T\}$ and $\mathbf{s}=\{m_{\chi},\bar{\sigma}_e\}$ are vectors containing background and signal related parameters, respectively. We neglect the contributions for $n>$\qe{2}, with $\epsilon^i_{n>2}$ being $<10^{-21}$.

The background model, $B^i(n_q | \eta)=\text{Pois}(n_q|\eta)$, represents the Poisson probability of collecting $n_q$ charges in a pixel, given a mean number of dark current counts per pixel equal to $\eta$. We model $\eta$ as:
\begin{equation}
    \eta^{jl}(T_i) = \eta_{0} + c^{jl} + \alpha_{T}^jT_i,    
\end{equation}
where $\eta_{0}$
is a reference value and  $c^{jl}$ is the offset from  $\eta_{0}$ for CCD-$j$ and D2-$l$ segment. 

The signal model, $S^i(n_q|m_{\chi},\bar{\sigma}_e,t_i)$, gives the probability to collect $n_q$ charges in a pixel for DM particles with mass
$m_{\chi}$ and cross section $\bar{\sigma}_e$.
The time dependence of the signal accounts for the expected daily modulation. The procedure used to derive $S^i$ is described in the following.

The differential scattering rate for a DM-electron interaction depositing an energy $E_e$ with momentum transfer $q$ is given by \cite{Essig:2015cda}: 
\begin{equation}\label{eq:dR_dEe}
    \frac{dR}{dE_{e}} \propto \bar{\sigma}_{e}\int \frac{dq}{q^{2}} \left[ \int \frac{f(\mathbf{v},t_i)}{v} d^{3}v \right] \left| F_{\mathrm{DM}}(q) \right|^{2} \left| F_{c}(q,E_e)\right|^{2}\,,
\end{equation}
where $F_c$ is the convolution of in-medium screening effects and the crystal form factor, which encodes the electronic response of the target. For Hidden Sector particles, the DM form factor, $F_{\mathrm{DM}}(q)$, depends on the mass of the mediator, which we assume to be a dark photon (\mA). 
 For a heavy-mediator (\mA$\gg q$), we use $F_{\rm{DM}} = 1$. For an ultralight-mediator (\mA$\ll q$), we set $F_{\rm{DM}}= (\alpha m_e/q)^2$, where $m_e$ is the mass of the electron and $\alpha$ is the fine structure constant. 

The function $f(\mathbf{v},t_i)$ describes the velocity distribution of DM particles reaching the detector. Its time dependence arises from the Earth's motion through the Galaxy and, for sufficiently large DM–scattering cross sections, from the number of atoms 
in the Earth and atmosphere that the DM particles must traverse before reaching the detector.

The velocity distribution of the Galactic DM halo is assumed to follow a Maxwell-Boltzmann distribution in the Galactic frame. The velocity distribution at the detector is then obtained by boosting into the rest-frame of the laboratory, which is moving with velocity: 
\begin{equation}
\mathbf{v}_\mathrm{lab}(t) = \mathbf{v}_{\odot} + \mathbf{v}_{E}(t) + \mathbf{v}_\mathrm{rot}(t)\,,
\end{equation}
where $\mathbf{v}_{\odot}$  is the velocity of the Sun in the Galactic frame; $\mathbf{v}_{E}(t)$ is the Earth's velocity in the Solar frame; and $\mathbf{v}_\mathrm{rot}(t)$ is the rotational velocity of the laboratory in an Earth-centered inertial frame~\cite{McCabe:2013kea,Baxter:2021pqo}.
At a given time, the mean velocity of the DM flux (averaged over the entire velocity distribution) is given by $\langle \mathbf{v}_{\chi}\rangle = -\mathbf{v}_\mathrm{lab}$.

\begin{figure}
    \centering
    \includegraphics[width=\linewidth]{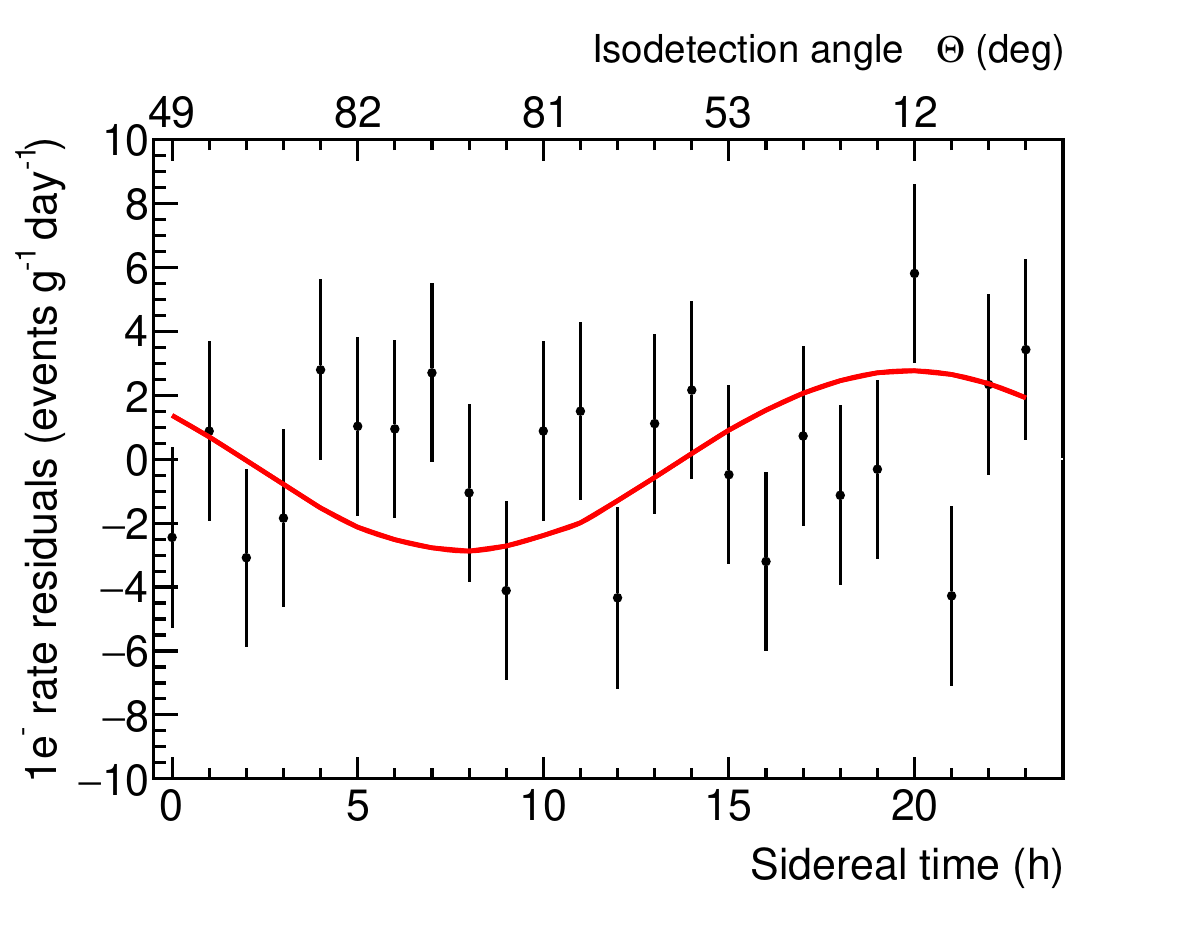}\\
    \caption{Residuals after the subtraction of the best-fit background-only model to the total rate, binned as a function of local apparent sidereal time. As a reference, the upper x-axis gives the isodetection angle $\Theta(t)$ for the first day of data taking. Each data point is obtained from the average of $~50$ images. The red line shows the expected signal (minus its time average) for a DM particle of mass $1$\MeV{}, $\bar{\sigma}_e=10^{-33}$cm$^{2}$ interacting via an ultralight dark photon mediator. 
    }
    \label{fig:rate_vs_sidereal}
\end{figure}

The velocity distribution also depends on the isodetection angle $\Theta(t_i)$, defined as the angle between the local zenith of the experiment and the mean incoming DM flux direction.
For $\Theta=0^{\circ}$, the mean DM flux comes from directly above the detector, while at $\Theta=180^{\circ}$, it comes from directly below.
As the Earth rotates, $\Theta$ varies over a sidereal day. 
The typical distance through the Earth that the DM must travel before reaching the detector varies with $\Theta(t_i)$, and consequently, so does the number of DM particles scattering in the Earth. 
This leads to a daily modulation in the expected DM event rate (see Fig.~\ref{fig:rate_vs_sidereal}). 
During the data-taking period, the isodetection angle at LSM (45.2$^{\circ}$ N) varied within the range $\Theta \in [3^{\circ},86^{\circ}]$. In this work, we approximate $\mathbf{v}_{E}(t)$ to its average over the data-taking period ($\langle\mathbf{v}_{E}\rangle=238.6$~km/s), rather than using the commonly adopted value~\cite{Baxter:2021pqo}. 

Several formalisms have been developed to model the DM flux incorporating the effects of Earth scattering (see, 
for example, ~Refs.~\cite{Kouvaris:2014lpa,Kavanagh:2016pyr,Davis:2017noy,Kavanagh:2017cru,Emken:2017qmp,Emken:2018run,Hooper:2018bfw,Bramante:2018qbc,Mahdawi:2018euy,Cappiello:2023hza}). In this work, we use the \texttt{VERNE 2.1} code~\cite{Verne}, which models the elastic interactions of light DM particles with Earth's nuclei\footnote{For DM particles at and above the MeV-scale, the DM-proton cross-section is typically much larger than the DM-electron cross-section~\cite{Emken:2019tni}. This means that the dominant Earth-scattering mechanism comes from recoils with nuclei (even though these recoils would not be energetic enough to be detectable individually, hence the need to search for electron recoils in the CCD target).}, assuming that they travel in straight lines until they scatter. After scattering, DM particles may be reflected back along their initial trajectory, with this probability derived from the full distribution of scattering angles for a given DM-nucleus interaction. In this approximation, only scattering in the forward or backward directions is possible, meaning that deflection through intermediate angles is neglected. 

However, as shown in~\cite{Verne2}, this approach agrees with full 3D Monte Carlo simulations to within 10\% in the Northern Hemisphere, while offering significantly improved computational efficiency\footnote{See also Ref.~\cite{Bertou:2025adb} for a comparison between the different approaches.}.

We compute the differential DM scattering rate via Eq.~\eqref{eq:dR_dEe}, modeling the crystal properties with \qcdark{}~\cite{QCDark:2024}, which also incorporates screening through an analytical dielectric model. For comparison with previous results, we also perform the analysis using \qedark{}~\cite{QEDark:2016}, where screening corrections are not taken into account. 

The number of electron--hole pairs, $n^{\prime}_q$, produced by a given deposited energy $E_e$, is calculated using the probability distribution $\mathcal{P}_{_{eh}}(n^{\prime}_q|E_e)$, which was derived in Ref.~\cite{Ramanathan:2020} using a semi-empirical approach.  

Finally, the probability of measuring in a pixel a number of $n_q$ electrons following DM interactions is given by:
\begin{equation}
S^{i}({n_q}|\mathbf{s},t_i)\propto\sum\limits_{n_q^{\prime}} \mathcal{P}(n_q|n^{\prime}_q) \int \frac{dR}{dE_e}\mathcal{P}_{_{eh}}(n_q^{\prime}|E_e)\, dE_e\,, 
\end{equation}
where $\mathcal{P}(n_q|n^{\prime}_q)$ denotes the probability of collecting a charge $n_q$ in a pixel from $n^{\prime}_q$ electrons-hole pairs that were generated, thereby accounting for the detector response. This probability is obtained from Monte Carlo simulations of the charge diffusion process, with $\sigma_{xy}^2$ calibrated using cosmic-ray data from a surface laboratory. The simulations  also incorporate the $100\times1$ pixel binning used during data-taking.

We fit the data by maximizing a joint likelihood function evaluated at 14 fixed DM masses between 0.53 and 2 $\rm{MeV/c^2}$.
The joint likelihood is defined as the product of the individual CCD likelihoods. For a given dark matter mass, the likelihood for a single CCD is computed as:

\begin{equation}\label{eq:binned_likelihood}
 \mathcal{L}(\bar{\sigma}_e;\mathbf{b})=\prod_{i=1}^{N_\mathrm{img}} \frac{{F_1^i}^{N_1^i}e^{-F_1^i}}{N_1^i!}\,.
\end{equation}
In $\mathbf{b}$, we set $\eta_0$ to be free and fix $c^{jl}$ and $\alpha^j_T$ according to the results obtained in Sec.~\ref{sec:dma} and reported in Table~\ref{tab:alpha_t}. We validated this approach by letting all parameters free and finding the same results.
The fit finds no preference for a DM signal at any mass. Consequently, we derive the 90\% C.L. upper limits on the DM-electron interaction cross section. We follow the approach of Ref.~\cite{Cowan:2013} and use the profile likelihood ratio test
statistic:
\begin{equation}
    \tilde q_{\mu} = 
    \left\{ \! \! \begin{array}{ll}
    \;- 2 \ln \tilde \lambda(\mu)  & \hat{\mu} \le \mu  \;, \\*[0.2 cm]
    \;\;\;\;\;\; 0 & \hat{\mu} > \mu \;,
    \end{array}
    \right.
\end{equation}
where $\mu$ is the signal strength parameter, which in this case corresponds to the DM-electron scattering cross section, $\bar{\sigma}_e$. 
For a given hypothesized value of $\bar{\sigma}_e$ and fixed DM mass, we define $\tilde\lambda(\bar{\sigma}_e)$ as:
\begin{equation}
    \tilde\lambda(\bar{\sigma}_e)=
    \left\{ \! \! \begin{array}{ll}
    \;\frac{\mathcal{L}(\bar{\sigma}_e;  \mathbf{\Hat{\Hat{b}}})}{\mathcal{L}(0;\mathbf{\hat{b})}}\,& \hat{\mu} < 0  \;, \\*[0.2 cm]
    \;\frac{\mathcal{L}(\bar{\sigma}_e;  \mathbf{\Hat{\Hat{b}}})}{\mathcal{L}(\Hat{\bar{\sigma}}_e;\mathbf{\hat{b})}}\,& 0 \le \hat{\mu} \le \mu  \;,
    \end{array}
    \right.
\end{equation}
where $\mathbf{b}$ is tested as nuisance parameter and marginalized over. By definition, $\mathbf{\Hat{\Hat{b}}}$ denotes the value of $\mathbf{b}$ that maximizes the $\mathcal{L}$ for a fixed value of $\bar{\sigma}_e$. The estimators $\Hat{\bar{\sigma}}_e$ and $\mathbf{\Hat{b}}$ correspond to the values that maximize $\mathcal{L}$, subject to the constraint that $\Hat{\bar{\sigma}}_e$ is non-negative, reflecting the fact that the presence of the signal can only increase the event rate. 

We verify through Asimov datasets that the test statistics follow the expected $\frac{1}{2} \chi^{2}$ distribution and derive the 90\% C.L. exclusion limits on $\bar{\sigma}_e$ accordingly. 

\begin{figure*}[!htb]
    \centering
    \includegraphics[width=0.45\linewidth]{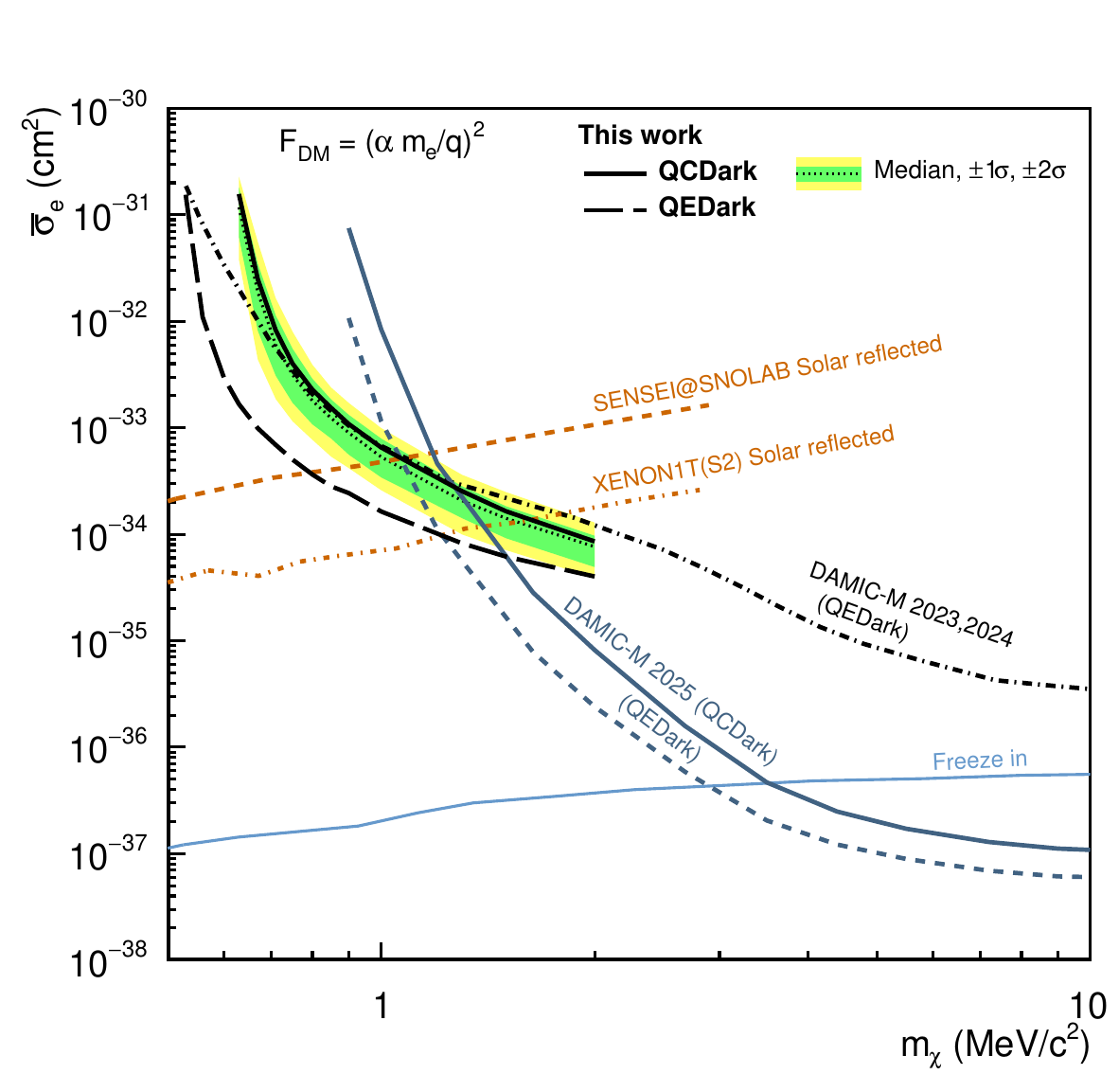}
    \hspace{0.6cm}
    \includegraphics[width=0.45\linewidth]{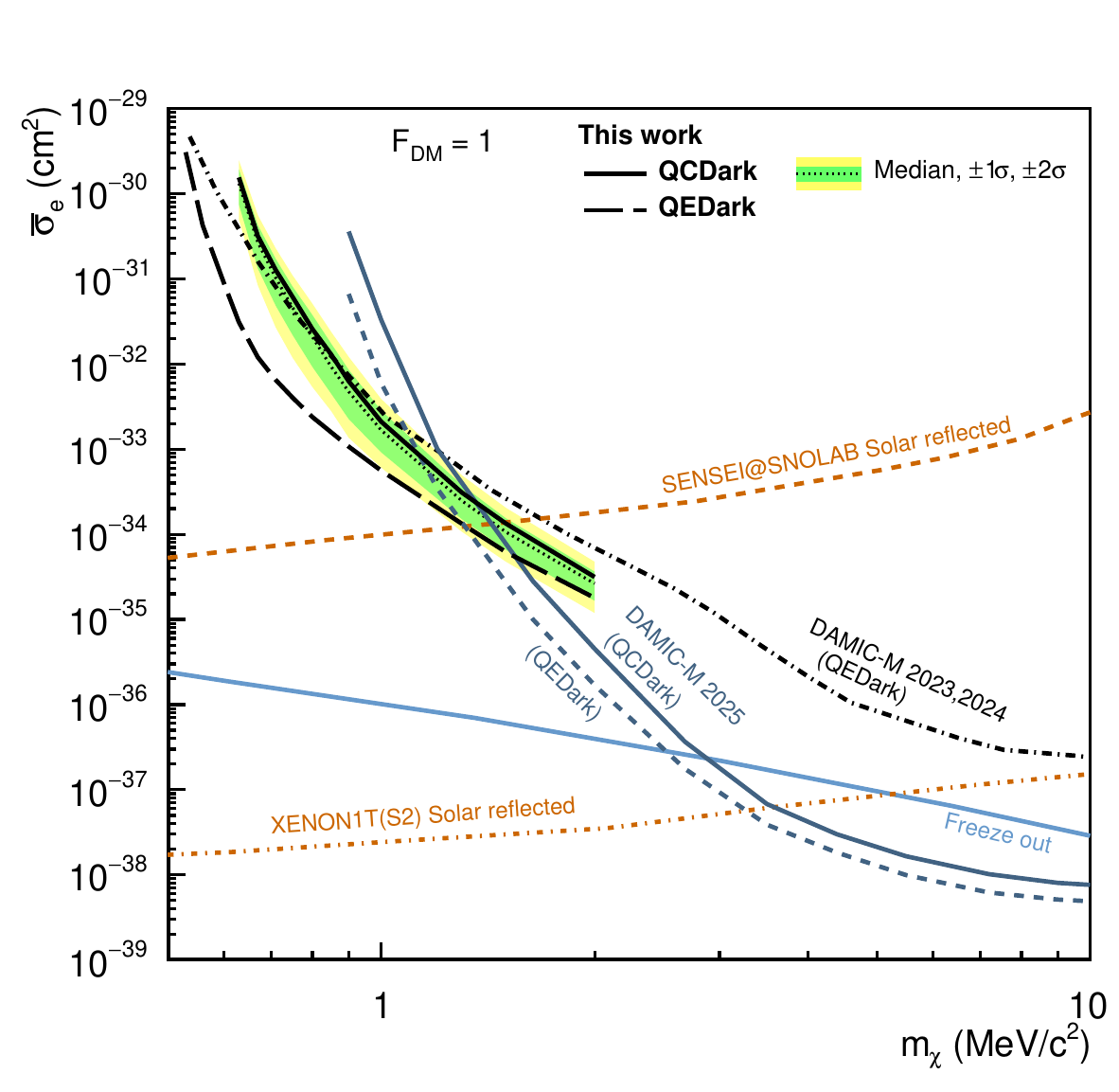}
    \caption{DAMIC-M 90\% C.L. upper limits (solid thick black) on DM-electron interactions through an ultralight (left) and heavy (right) dark photon mediator obtained from the daily modulation analysis. In green and yellow are the 1- and 2$\sigma$ sensitivity bands. The limits derived using the QEDark model are also reported (dashed thick black). 
    Also shown are previous DAMIC-M limits \cite{DAMIC-M:2023FC,DAMIC-M:2023DM,DAMIC-M:2025} for Galactic DM, as well as results from other experiments, such as SENSEI@SNOLAB \cite{SolarReflectionSensei} and XENON1T \cite{SolarReflectionXenon}, for solar-reflected DM. Theoretical expectations assuming a DM relic abundance from freeze-in and freeze-out mechanisms are also shown in light blue \cite{freezein}.}. 
    \label{fig:upper_limits}
\end{figure*}

The exclusion limits obtained for ultralight and heavy mediators are shown in Fig.~\ref{fig:upper_limits}. These limits fall within the expected 64\% sensitivity band estimated by the Asimov datasets (also shown in Fig.~\ref{fig:upper_limits}). Note that the modeled background is data-driven and it cannot be sampled with Monte Carlo simulations. Our Asimov datasets are thus constructed by breaking the time dependence of the data, and hence removing any possible modulation in a hypothetic signal. Consequently, systematic errors are not taken into account for the sensitivity bands. 
The daily modulation analysis presented here yields  a notable enhancement over the previous DAMIC-M constraints based on the same data~\cite{DAMIC-M:2025}, improving sensitivity by up to~2 orders of magnitude. Note that there is a maximum cross section that can be excluded since Earth scattering effects eventually lead to a complete attenuation of the DM flux. For the ultralight mediator, our exclusion limit is valid up to $\bar{\sigma}_e \approx 10^{-27}\,\mathrm{cm}^2$ across the explored mass range. For the heavy mediator, the limit is valid up to $\bar{\sigma}_e \approx 10^{-27}\,\mathrm{cm}^2$ for $m_{\chi}=1$\MeV, with the upper bound decreasing with increasing mass to $\bar{\sigma}_e \approx 10^{-30}\,\mathrm{cm}^2$ for $m_{\chi}=10$\MeV{}~\cite{Emken:2019tni}.

In this analysis we focused on single-electron events, which dominate the expected signal for low DM masses ($\leq 2.7$\MeV{}). As $m_{\chi}$ increases, the fraction of two-electron events becomes significant. However, the predicted modulation amplitude of the two-electron signal for cross sections already excluded by previous DAMIC-M constraints remain below the detection threshold given the available statistics. For this reason, we do not extend the modulation analysis beyond the single-electron rate.


\section{Discussion}

We present a dedicated search for a daily modulation signal from Hidden Sector DM particles in the Galactic halo with an improved dataset with larger exposure and lower backgrounds from the DAMIC-M prototype detector. 
The characteristic daily modulation of the single-electron rate, caused by the scattering of DM particles with sufficiently large cross section as they travel through the Earth, offers a powerful means of distinguishing potential low-mass DM signals from dark current backgrounds.

Our model-independent analysis finds no statistically significant periodic signals in the explored frequency range, including the sidereal frequency where a dark matter signal would be expected. This result demonstrates the excellent stability of the detector. 

We performed a model-dependent analysis targeting the predicted daily modulation signal expected from Hidden Sector particles interacting with electrons. 
Within the explored mass range, we find no statistically significant preference for a DM signal. As a result, we set new and stringent 90\% C.L. exclusion limits on the DM–electron scattering cross section.

The exclusion limits (Fig.~\ref{fig:upper_limits}) are derived using both the \qedark{}~\cite{QEDark:2016} and \qcdark{}~\cite{QCDark:2024} models. 
Our exclusion limits surpass all previously published results for DM particles from the Galactic Halo in the mass range 0.53–1.2\MeV, considering both heavy and light mediator scenarios.
Results from the \qcdark{} model, although less stringent than those from \qedark{}, are based on a more accurate and up-to-date description of DM–electron interactions, particularly at low energies.

With this dedicated search for a daily modulating DM signal, the DAMIC-M prototype detector has established, yet again, world-leading constraints on the electron interactions of DM particles with MeV-scale masses. 
This work highlights the strong potential of the DAMIC-M experiment for future DM searches.


\section{Acknowledgements}
We would like to thank the Modane Underground Laboratory and its staff for support through underground space and logistical and technical services. LSM operations are supported by the CNRS, with underground access facilitated by the Soci\'et\'e Fran\c caise du Tunnel Routier du Fr\'ejus. The DAMIC-M project has received funding from the European Research Council (ERC) under the European Union’s Horizon 2020 research and innovation program Grant Agreement No. 788137, and from the NSF through Grant No. NSF PHY-1812654. The work at the University of Chicago and University of Washington was supported through Grant No. NSF PHY-2413013 and 2413014. This work was supported by the Kavli Institute for Cosmological Physics at the University of Chicago through an endowment from the Kavli Foundation. We also thank the Krieger School of Arts \& Sciences at Johns Hopkins University for its contributions to the DAMIC-M experiment. The IFCA was supported by Project  DMPHENO2LAB (PID2022-139494NB-I00) financed by MCIN/AEI/ 10.13039/501100011033/FEDER, EU. N.C.-M. acknowledges funding from the Ramón y Cajal Grant RYC2022-038402-I, financed by MCIN/AEI/10.13039/501100011033 and the FSE+. The University of Z\"{u}rich was supported by the Swiss National Science Foundation. We thank the Boulby Underground Laboratory, SNOLAB, Canfranc Underground Laboratory, Washington Nanofabrication Facility and their staff for support through underground space, logistical and technical services. The CCD development work at Lawrence Berkeley National Laboratory MicroSystems Lab was supported in part by the Director, Office of Science, of the U.S. Department of Energy under Contract No. DE-AC02-05CH11231. We thank Teledyne DALSA Semiconductor for CCD fabrication.

\bibliography{references.bib}

\end{document}